# Influence of magnetism on the superconducting properties of iron-based superconductor NdFeAsO$_{0.88}$F$_{0.12}$

Y. Ding[1], Y. Sun[1], X. D. Wang[1], H. C. Wang[1], Z. X. Shi[1], Z. A. Ren[2], J. Yang[2], W. Lu[2]

Corresponding author Z. X. Shi: zxshi@seu.edu.cn

1.Department of Physics, Southeast University, Nanjing, China 211189

Mail address: dingyiseu@gmail.com

2.Institute of Physics and Beijing National Laboratory for Condensed Matter Physics, Chinese Academy of Sciences, Beijing 100190, P. R. China

Abstract

Both DC and AC magnetization measurements were performed on the NdFeAsO$_{0.88}$F$_{0.12}$ superconductor to investigate the influence of magnetism on the superconducting properties of this system. The crossover of the ZFC and FC magnetic susceptibility $\chi(T)$ curves under 7.5KOe was observed. The imaginary component of the first harmonics of the AC magnetic susceptibility, $\chi''(T)$, increases with the increasing DC field $H_{dc}$ below 10K and shows frequency dependency under 7.5KOe at low temperature. The paramagnetism of $Nd^{3+}$ ions tilts the magnetic hysteresis loops and broadens the hysteresis width $\Delta M$. After correction for the paramagnetism, the field and temperature dependence of intrinsic $J_{cm}$ was obtained and compared with the experimentally obtained total $J_{tot}$. The origin of the abnormal behavior of $\chi(T)$ and $\chi''(T)$ was investigated and attributed to a magnetic background, which was speculated to be caused by the spin-glass state. However, this magnetic background does not affect the flux pinning properties in this sample. The related mechanism was discussed.

## 1. Introduction

The recent discovery of superconductivity at 26K in the iron oxypnictide LaFeAs(O, F) [1] has stimulated great interests among condensed-matter physics community. Tremendous work was carried out, leading to the emergence of new iron-based superconductor families with different crystal structures: 1111 (REFeAs(O, F)), 122 ((Ba, K)Fe$_2$As$_2$) [2], 111 (LiFeAs) [3] and 11 (Fe(Se, Te)) [4]. The REFeAs(O, F) superconductors, which has the highest $T_c$ when La was replaced by Sm [5], crystallize in the tetragonal P4/nmm space group with two formula units per unit cell. The crystal structure consists of alternating RE-O and Fe-As layers stacked along the c axis. Since magnetic elements such as Fe and rare earth exist in this system, relevant questions have been raised about the relationship between magnetism and superconductivity. Neutron, muon spin rotation ($\mu SR$) and Mossbauer measurements [6, 7] on undoped LaFeAsO have revealed commensurate spin density wave (SDW) order of the Fe moments below $T_N$=135K with amplitude of 0.35$\mu_B$. Superconductivity can be induced from the magnetically ordered parent compound by carrier doping, external and internal pressure. Some measurements suggest that the magnetic order is rapidly suppressed upon doping and the maximum $T_c$ is achieved just as static magnetism disappears [6, 8]. Tarantini et al excluded the long-range antiferromagnetic order in the superconducting NdFeAsO$_{0.94}$F$_{0.06}$ [9]. However, Drew et al detected magnetic fluctuations [10] and static magnetism [11] coexisting in the SmFeAsO$_{1-x}$F$_x$ by muon spin rotation, and Ryan et al detected the coexistence of long-ranged magnetic order and superconductivity in SmFeAsO$_{1-x}$F$_x$ by using neutron diffraction [12]. The coexistence of magnetism and superconductivity, and the role played by magnetism in the basic superconducting mechanism need to be further investigated both theoretically and experimentally.

Besides the studies related to the superconducting mechanism, it is also necessary to study the influence of magnetism on superconductivity. Previous reports [9, 13, 14] have shown the modified electromagnetic behavior by magnetism in the 1111 system. In this paper, both DC and AC magnetization measurements were performed. The influence of magnetism background on the superconducting properties is investigated systematically.

## 2. Experimental

The superconducting NdFeAsO$_{1-x}$F$_x$ sample was prepared directly by a high-pressure (HP) method similar to reference [15]. The starting materials were mixed according to the nominal stoichiometric ratio of NdFeAsO$_{0.88}$F$_{0.12}$, then ground thoroughly and pressed into pellets. The pellets were sealed in boron nitride crucibles and sintered in a high pressure synthesis apparatus under 6Gpa at the temperature of 1250℃ for 2 hours. The HP sample has the dimensions of $1.16 \times 1.88 \times 4.50$ mm$^3$, weights 58mg and is about 82% of the theoretical density (7.21g/cm3). The transition temperature $T_c$ determined by resistivity measurements (not shown here) is about 51K and the resistivity $\rho(T_c)$ is about 2.04mΩcm.

The crystal structure of HP sample was characterized by powder X-ray diffraction (XRD) on an MXP18A-HF–type diffractometer with Cu-Kα radiation from 20◦ to 80◦ with a step of 0.01◦. As shown in Figure 1, all main peaks can be well indexed based on the ZrCuSiAs tetragonal structure, confirming the formation of NdFeAs(O, F) phase. No impurity peak was observed.

DC magnetization $M(T)$ and $M(H)$ measurements were performed using a vibrating sample magnetometer (VSM) of a Quantum Design Model 6000 PPMS. Magnetic critical current densities $J_{cm}$ were estimated using the Bean model from the *M(H)* hysteresis loops (MHLs).By using the same PPMS equipped with an AC susceptometer, the first harmonics of the AC magnetic susceptibility as a function of the temperature were measured at various frequencies $\nu$, AC magnetic field amplitudes $h_{ac}$ and DC fields $H_{dc}$, which was parallel to the AC field. Before each measurement, the sample was warmed up to 60K to fully expel the flux trapped inside. Sufficient waiting time was adopted to ensure the thermal homogeneity. During both DC and AC measurements, the magnetic fields were applied along the longest dimension of the sample.

## 3. Results and discussion

*3.1* The DC magnetic susceptibilities

The DC magnetic susceptibilities $\chi(T)$ for HP sample were measured by zero-field cooling (ZFC) and field cooling (FC) under 15Oe, 5KOe, 7.5KOe and 15KOe. The results are plotted in Figure 2. Figure 2 (a) shows the $\chi(T)$ curves measured under 15Oe. The sharp diamagnetic superconducting transition indicates the good sample quality. The superconducting shielding volume fraction is about 99%, confirming bulk superconductivity in this sample. The transition temperature $T_c$ determined by the onset of diamagnetic signal is about 50K. Figure 2 (b) shows the $\chi(T)$ curves measured under 5KOe. The magnetic fields suppressed the superconducting diamagnetism, and a broad peak emerged in the ZFC curve at about 11K. This peak was also detected in the $\chi(T)$ curve measured under 7.5KOe as shown in Figure 2 (c), where the ZFC curve became positive in the whole temperature range due to the strong paramagnetic background. Similar behavior was reported by several groups in Nd [9], Sm [14] and Pr [16] 1111 superconductors. What to our surprise is, the $\chi(T)$ shows an abnormal behavior that the ZFC curve went above FC curve between 10K and 25K. Further increase of the magnetic field to 15KOe results in the increase of the paramagnetic component and eliminated the peak. The ZFC curve, however, was still located above the FC curve in the temperature range of 5 to 30K. This phenomenon so far to our knowledge has not been reported in the iron-based 1111 system.

To investigate the origin of this behavior, the paramagnetic background was evaluated and subtracted from the total susceptibilities. Assuming that the paramagnetic magnetic moments of the $Nd^{3+}$ ions do not interact with each other, the paramagnetic magnetization is expected to be given by

$$M_{pm} = M_0 B_J(g\mu_B JB / k_B(T+\Theta)) \tag{1}$$

where $g$ is the Lande factor, $\mu_B$ is the Bohr magneton, $k_B$ is the Boltzmann constant, $B_J(x)$ is the Brillouin function and $M_0 = NgJ\mu_B$ (N is the number of Nd ions per unit volume). A phenomenological parameter $\Theta$ was introduced because the electric crystal field (CEF) acting on Nd 4f electrons splits and mixes the degenerated $|JM_J\rangle$ states resulting in, for $g\mu_B JB/k_B T \ll 1$, the modification of the Curie form $\chi = C/T$ to the Curie-Weiss law, $\chi=C/(T+\Theta)$ [17]. According to the Russel-Saunders coupling model the ground state term of the $Nd^{3+}$ ion is $^4I_{9/2}$ ($J=9/2$).

The $M(T)$ curve measured from 50 to 300K under 7.5KOe was plotted in Figure 3 (a), showing the Curie-Weiss behavior similar to [18]. The paramagnetic background of total magnetization from 5 to 50K was obtained by fitting the measured $M(T)$ curve using equation

(1) with parameter $\Theta$ =16K and $M_0V$=3.26emu, where $V$ is the sample volume. After subtracting the paramagnetic background, the $\chi(T)$ curves under 5KOe, 7.5KOe and 15KOe were shown in Figure 3 (b)-(d). The $\chi(T)$ curves under 5KOe and 7.5KOe show negative values below the superconducting onset temperature, indicating the diamagnetic effect. Besides, four features can be clearly noted: (1) The FC $\chi(T)$ curves in Figure 3 (b)-(d) raise up below 15K. This may be caused by the underestimating of the paramagnetic component at low temperature, where the $M_{pm}$ value deviates from equation (1). (2) The ZFC $\chi(T)$ curves under 5KOe and 7.5KOe drop dramatically at low temperatures as shown in Figure 3 (b) and (c). In a pervious report [16], this was explained by the onset of intergrain superconductivity. (3) The ZFC $\chi(T)$ curve shows a broad peak around 16K in Figure 3 (c). (4) Part of the FC $\chi(T)$ curve under 5KOe and 7.5KOe goes below ZFC curve shown in Figure 3 (c) and (d). As for feature (3) and (4), Tarantini et al suggested the possibility of spin glass in the Nd-1111 system [9]. Indeed, similar complex $\chi(T)$ behavior was observed on some glassy system on a much smaller field scale [19, 20], indicating different ordering and/or freezing processes during ZFC and FC. In order to evaluate this possibility, AC magnetic susceptibilities measurements were performed.

*3.2* The AC magnetic susceptibilities

Figure 4 (a) shows the AC magnetic susceptibilities $\chi'$ and $\chi''$ as a function of temperature at frequencies of 100Hz, 500Hz and 1kHz, for DC magnetic fields $H_{dc}$=7.5KOe and the AC magnetic field $h_{ac}$=8Oe. As the frequency decreases, the transition temperature shifts to lower value, the peak height in $\chi''$ decreased as it moves to lower temperature.

Figure 4 (b) shows the $\chi'(T)$ and $\chi''(T)$ curves measured at the fixed frequency $\nu$ =500Hz, $h_{ac}$=1Oe and at various $H_{dc}$=50Oe, 150Oe, 500Oe, 5KOe and 7.5KOe. As the dc fields increase, the transition in $\chi'(T)$ and $\chi''(T)$ becomes broad and the peak height in $\chi''$ decreased as it moves to lower temperature. These features are similar to that of high-temperature superconductors [21, 22] and can be explained by flux creep and the temperature dependence of shielding currents [23]. In Figure 4 (a), however, $\chi''$ was frequency independent in the temperature range of 25K to 30K but frequency dependent from 5 to 25K. This cannot be explained by flux creeping, which is weaker in lower temperatures. Moreover, it is also noted that an abnormal behavior of $\chi''$ evolves with increasing $H_{dc}$ in Figure 4 (b). A weak peak in $\chi''$ at about 12K emerges when $H_{dc}$=500Oe, and the peak seems to move below 5K when $H_{dc}$ reaches 5KOe. One may attribute this behavior to the superconducting intergranular component, corresponding to the drop of $\chi$ value at about 10K in Figure 3 (b) and (c). However, recent report [24] investigated the granularity in LaFeAsO$_{0.92}$F$_{0.08}$, showing that the intergranular component is much larger than the intragranular one. On the other hand, the peak in $\chi''$ caused by the intergranular weak links can only be observed under low DC fields much smaller than 5KOe. Thus, the dissipation peak and the drop in dc $\chi$ value below 10K are not likely caused by intergranular coupling at low temperatures, unless there exists small amount of 'medium links', which can survive under more than 5KOe. We suggest that the abnormal behavior of $\chi''$ may be caused by the spin glass, which shows similar frequency dependence as $\chi''$ in Figure 4 (a) [25]. To fully investigate this possibility, further study in details is under process.

*3.3*  Magnetic hysteresis loops

Figure 5 (a) shows the magnetic hysteresis loops (MHLs) of the HP sample up to 1.5T at several temperatures from 5 to 25K. The MHLs shows a general behavior of superconductors that the hysteresis loop width $\Delta M$ decreases with increasing temperature. However, the MHLs were shifted to higher value by a paramagnetic background caused by $Nd^{3+}$ ions, resulting in a crossover of MHLs at different temperatures in the field range of 5KOe to 1T. This behavior was also observed in other reports [13, 18]. The experimentally measured magnetization contains two components: irreversible superconducting magnetization $M_{sc}$ and paramagnetic magnetization $M_{pm}$ of $Nd^{3+}$ ion.

$$M = M_{sc} + M_{pm} \qquad (2)$$

The paramagnetic magnetization $M(H)$ not masked by superconductivity was measured at 53K up to 9T as shown in the inset of Figure 5 (a). $M_0V$=2.55emu of the sample was determined by fitting the $M(H, 53K)$ curve using equation (1) with $\Theta=0$ for simplicity. The paramagnetic background of total MHLs measured at 5 to 25K was then calculated using equation (1) with parameter $\Theta$ =7K to 9K. After subtracting $M_{pm}$ from measured MHLs, the superconducting MHLs were obtained and shown in Figure 4 (b). After the paramagnetic correction the MHLs show no abnormal from 5KOe to 1T, so it seems that the crossover of MHLs is simply caused by a superposition of paramagnetic background on the superconducting irreversible magnetization. The magnetic background related to the possible spin glass and/or magnetic impurities have no contributions to the flux pinning property, thus do not affect the MHLs.

It is noted that a centre peak appears in the MHLs at each temperature. Similar zero-field peak was also observed in NdFeAsO$_{1-\delta}$ system reported by Moore et al [26] and hight temperature superconductors [27]. This centre peak is speculated to be caused by the existence of the Bean-Livingston surface barrier [28], and will be investigated in detail elsewhere.

*3.4 Effect of paramagnetism on $J_c$*

Because the distribution of the flux density in the sample during the increasing $B^-$ and during the decreasing $B^+$ field process at the same applied field is different, the paramagnetic magnetization during the increasing field process $M_{pm}^-$ and during the decreasing field process $M_{pm}^+$ should be different as well. The Bean critical state model leads to the following expressions for the magnetization in the decreasing $M^+$ and in the increasing $M^-$ field branches:

$$M^+ = \Delta M_{sc}/2 + \langle M_{pm}^+ \rangle \quad (3)$$

$$M^- = \Delta M_{sc}/2 + \langle M_{pm}^- \rangle \quad (4)$$

where $\Delta M_{sc} = M_{sc}^+ - M_{sc}^-$ is the magnetic hysteresis if there is no paramagnetic contribution, and $\langle M_{pm}^\pm \rangle$ is the spatial average of $M_{pm}^\pm$, respectively. From Equation (3) and (4) the width of the experimentally measured magnetic hysteresis $\Delta M$ should be:

$$\Delta M = \Delta M_{sc} + \Delta M_{pm} \quad (5)$$

where $\Delta M_{pm} = <M_{pm}^+> - <M_{pm}^->$. Considering a fully penetrated infinitely long slab of thickness $d$ along the $x$ axis and infinite in the $yz$ plane with the magnetic field $B$ parallel to the $z$ axis, and using the simple Bean critical state model with Equation (2), the field dependence of $\Delta M_{pm}$ can be deduced as [29]:

$$\Delta M_{pm}(H) = <M_{pm}^+> - <M_{pm}^-> = M_0(<B_J(\eta^+)> - <B_J(\eta^-)>) \quad (6)$$

where $\eta = g\mu_B B^{\pm}/k_B(T+\Theta)$, $B^-$ and $B^+$ is the flux density in the sample during the increasing field and during the decreasing field process respectively, and $<B_J(\eta^{\pm})>$ is the spatial average of $B_J(\eta^{\pm})$.

As can be seen from Equation (5), the intrinsic magnetic hysteresis $\Delta M_{sc}$ was broadened by paramagnetism to the experimentally measured (total) magnetic hysteresis $\Delta M$. Therefore, total magnetization critical current density $J_{tot} = 20\Delta M/a(1-a/3b)$ deviates from the intrinsic critical current density $J_{cm} = 20\Delta M_{sc}/a(1-a/3b)$, where $a$ and $b$ are thickness and width of the sample. The field dependence of the total $J_{tot}$ and intrinsic $J_{cm}$ at 5, 10, 15, 20 and 25K were plotted in Figure 5 (a) for comparison. $\Delta J_{cm} = J_{tot} - J_{cm}$ decreases monotonically with applied field as shown in the inset of Figure 5 (a). Figure 5 (b) shows the temperature dependence of $J_{tot}$ and $J_{cm}$ at 2KOe, 5KOe, 7.5KOe and 12KOe. The inset of Figure 5 (b) shows that $\Delta J_{cm}$ decreases quickly with increasing temperature. In Figure 5, it is obvious that $J_{tot}$ is larger than $J_{cm}$ at 5K. However in high field and high temperature, $\Delta J_{cm}$ can be neglected, the intrinsic critical current density $J_{cm}$ can be well approximated by the total critical current density $J_{tot}$. The reason is that in the high field and high temperature region, magnetic hysteresis decreases quickly due to weaken pinning, so the distribution of the flux density in the sample during the increasing $B^-$ and during the decreasing $B^+$ field process at the same applied field is very close, resulting in much less difference between $M_{pm}^-$ and $M_{pm}^+$. It is worth mention that the magnetic background shows no influence on $J_{cm}$ in whole temperature range as shown in Figure 6 (b), confirming that the magnetic background has no contribution to the flux pinning property in this sample.

The possible reason is that the magnetic background is an intrinsic one distributing homogeneously among the whole sample.

## 4. Conclusions

In summary, the influence of magnetism on the superconducting properties of iron-based superconductor NdFeAsO$_{0.88}$F$_{0.12}$ was studied by means of both DC and AC magnetization measurements. Our results lead to following conclusions: (1) The anormal behavior of $\chi(T)$ curves under 7.5KOe, the ac field frequency dependency of $\chi''(T)$ around 10K and the increasing of dissipation with dc field $H_{dc}$ at low temperatures suggest the possibility of spin glass existing in this system. (2) The paramagnetism of Nd$^{3+}$ ions tiled the magnetic hysteresis loops and broadened the hysteresis width, therefore the magnetization critical current $J_{cm}$ based on the Bean critical state model deviated from the intrinsic value. The experimental critical current density $J_{tot}$ derived from MHLs is larger than the intrinsic critical current density $J_{cm}$ at low temperatures. In high field and high temperature, the magnetism influence can be neglected, and $J_{cm}$ can be well approximated by $J_{tot}$. (3) The flux pinning property in this sample was not affected by the observed magnetic background, suggesting the background is intrinsic of this system.

**Acknowledge**

This work was supported by the scientific research foundation of graduate school of Southeast University (Grant No. YBJJ0933), by the Natural Science Foundation of China (NSFC, Grant No. 10804127), by the Cyanine-Project Foundation of Jiangsu Province of China (Grant No. 1107020060) and by the Foundation for Climax Talents Plan in Six-Big Fields of Jiangsu Province of China (Grant No. 1107020070).

**Reference**

[1] Kamihara Y, Watanabe T, Hirano M and Hosono H 2008 *J. Am. Chem. Soc.* **130**, 3296
[2] Rotter M, Tegel M, Johrendt D, Schellenberg I, HermesW and Pottgen R 2008 *Phys. Rev. B* **78** 020503
[3] Wang X C, Liu Q, Lv Y, GaoW, Yang L X, Yu R C, Li F Y and Jin C 2008 *Solid State Commun.* **148** 538
[4] Hsu F C *et al* 2008 *Proc. Natl Acad. Sci. USA* **105** 14262
[5] Ren Z A et al 2008 *Chin. Phys. Lett.* **25**, 2215
[6] Cruz C et al. 2008 *Nature* **453**, 899-902
[7] Klauss H H et al. 2008 *Phys. Rev. Lett.* **101**, 077005
[8] Liu R H et al. 2008 *Phys. Rev. Lett.* **101**, 087001
[9] Tarantini C, Gurevich A, Larbalestier D C, Ren Z, Dong X L, Lu W, Zhao Z X, 2008 *Phys. Rev. B* **78**, 184501
[10] Drew A J et al 2008 Phys. Rev. Lett. 101, 097010
[11] Drew A J et al 2009 Nature Materials 8, 310-314
[12] Ryan D H, Cadogan J M, Ritter C, Canepa F, Palenzona A, Putti M 2009 *Phys. Rev. B*, **80** 220503(R)
[13] Pissas M, Stamopoulos D, Ren Z A, Shen X L, Yang J Zhao Z X 2009 *Supercond. Sci. Technol.* **22** 055008
[14] Cimberle M R, Canepa F, Ferretti M, Martinelli A, Palenzona A, Siri A S, Tarantini C, Tropeano M, Ferdeghini C 2009 *J. Magn. Magn. Mater.* **321**, 3024-3030
[15] Ren Z A et al 2008 *Europhys. Lett.* **82**, 57002
[16] Bhoi D, Mandal P, Choudhury P, Dash S, Banerjee A 2010 arXiv: 1002.0208v1
[17] Penney W G, Schlapp R 1932 *Phys. Rev.* **41** 194
[18] Awana V P S, Meena R S, Pal A, Rao K V, Kishan H 2010 arXiv: 1003.0273v1
[19] Dahamni A, Taibi M, Nogues M, Aride J, Loudghiri E, Belayachi A 2002 *Mater. Chem. Phys.* **77** 912-917
[20] Kundu A K, Nordblad P, Rao C N R 2006 *J. Phys: Condes. Matter* **18** 4809-4818
[21] Gomory F, Takacs 1993 *Physica C* **217** 297
[22] Lee C Y, Kao Y H, 1995 *Physica C* **241** 167-180
[23] Qin M J, Yao X X, 1996 *Phys. Rev. B* **54** 10
[24] Phlichetti M, Adesso M, Zola D, Lou J, Chen G F, Zheng L, Wang N L, Noce C and Pace S 2008 *Phys. Rev. B* **78**, 224523
[25] Huang S L, Ruan K Q, Lv Z M, Wu H Y, Pang Z Q, Cao L Z, Li X G 2006 *J. Phys.: Condens. Matter* **18** 7135-7144
[26] Moore J D et al 2008 *Supercond. Sci. Technol.* **21** 092004
[27] Li P J, Wang Z H, Hu A M, Bai Z, Qiu L, Gao J 2006 *Supercond. Sci. Technol.* **19** 825-829
[28] Bean C P, Livingston J D, 1964 *Phys. Rev. Lett.* **12** 14-16
[29] Qin M J, Shi Z X, Ji H L, Jin X, Yao X X 1995 *J. Appl. Phys.* **78**, No. 5, 1 September

**Figure Captions**

Figure 1. X-ray powder diffraction patterns of the HP NdFeAsO$_{0.88}$F$_{0.12}$ sample.

Figure 2. ZFC and FC M(T) curves of the HP sample at external field (a) 15Oe, (b) 5KOe, (c) 7.5KOe, (d) 15KOe.

Figure 3. (a) Dash line indicates the *M(T)* curve measured from 50 to 300K under 7.5KOe. Solid line represents the paramagnetic magnetization calculated from equation (1). (b), (c), (d) ZFC and FC *M(T)* curves after subtracting the paramagnetic component at external field 5KOe, 7.5KOe, 15KOe, respectively.

Figure 4. (a) Magnetic hysteresis loops of the HP sample measured at 5, 7, 8, 9, 10, 11, 12, 13, 15, 20 and 25K. Inset shows the *M(H)* curve measured at 53K up to 9T. (b) Magnetic hysteresis loops after subtracting the paramagnetic background at 5-25K.

Figure 5. (a) AC magnetic susceptibilities $\chi'$ and $\chi''$ as a function of temperature at frequencies of 100Hz, 500Hz and 1kHz, for $H_{dc}$=7.5KOe and $h_{ac}$=8Oe. (b) The $\chi'(T)$ and $\chi''(T)$ curves measured at the fixed frequency $\nu$=500Hz, $h_{ac}$=16Oe and at various $H_{dc}$=50Oe, 150Oe, 500Oe, 5KOe and 7.5KOe.

Figure 6. (a) Comparison of $J_{tot}(H)$ and $J_{cm}(H)$ up to 1.4T at the temperature of 5, 10, 15 20 and 25K, where $J_{tot}(H)$ was extracted from MHLs in Figure 3 and $J_{cm}(H)$ is after paramagnetic correction. Inset shows the field dependence of $\Delta J_c = J_{tot} - J_{cm}$ at 5, 10, 15 20 and 25K. (b) Comparison of $J_{tot}(H)$ and $J_{cm}(H)$ in the temperature range of 5-25K at 2KOe, 5KOe, 7.5KOe and 12KOe. Inset shows the temperature dependence of $\Delta J_c = J_{tot} - J_{cm}$ in the temperature range of 5-25K at 2KOe, 5KOe, 7.5KOe and 12KOe

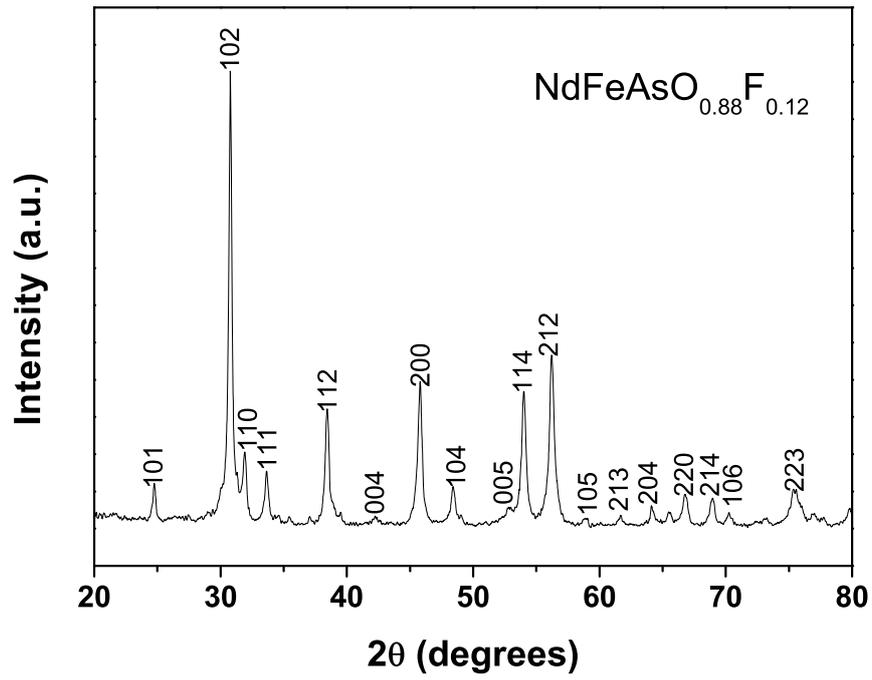

**Figure 1 Y. Ding** *et al.*

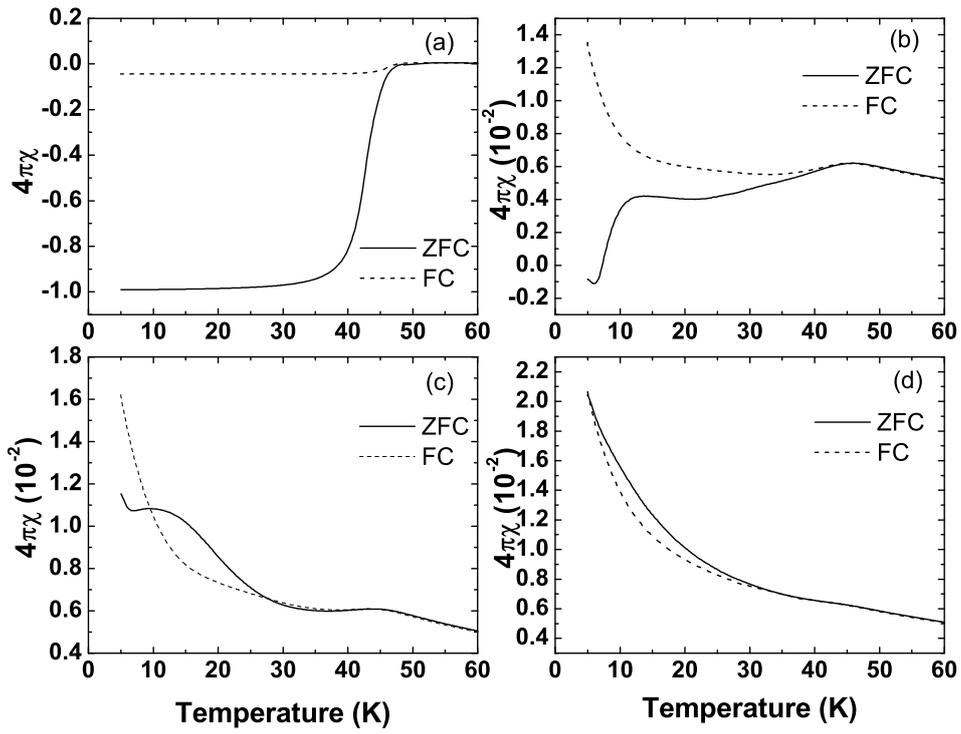

**Figure 2 Y. Ding** *et al.*

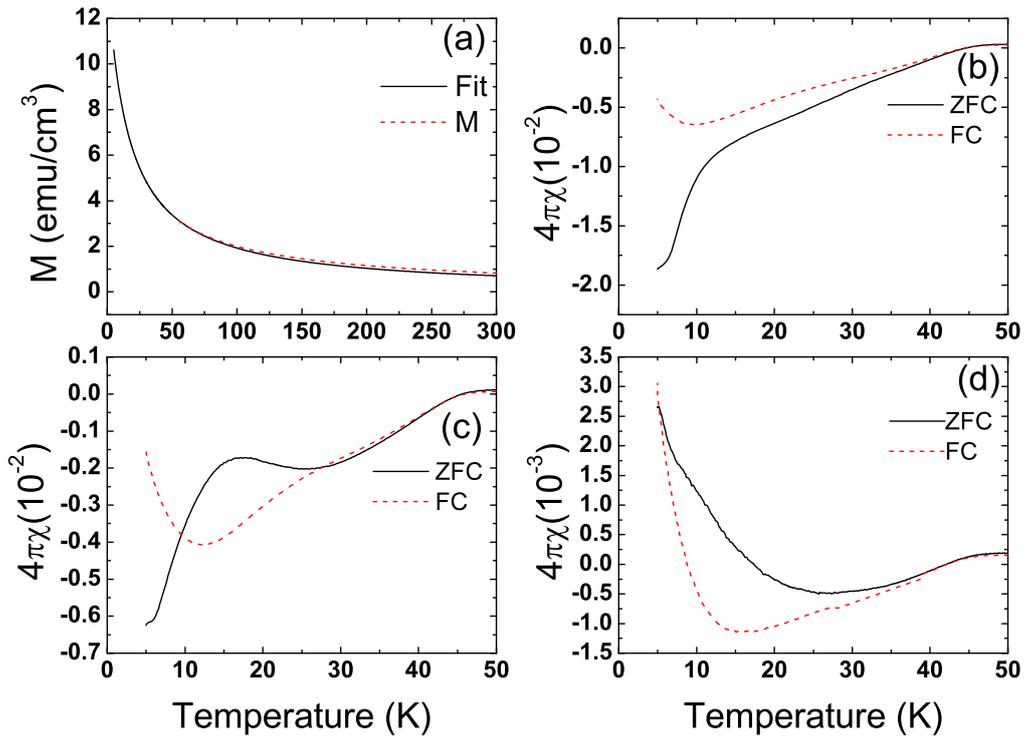

**Figure 3 Y. Ding *et al.***

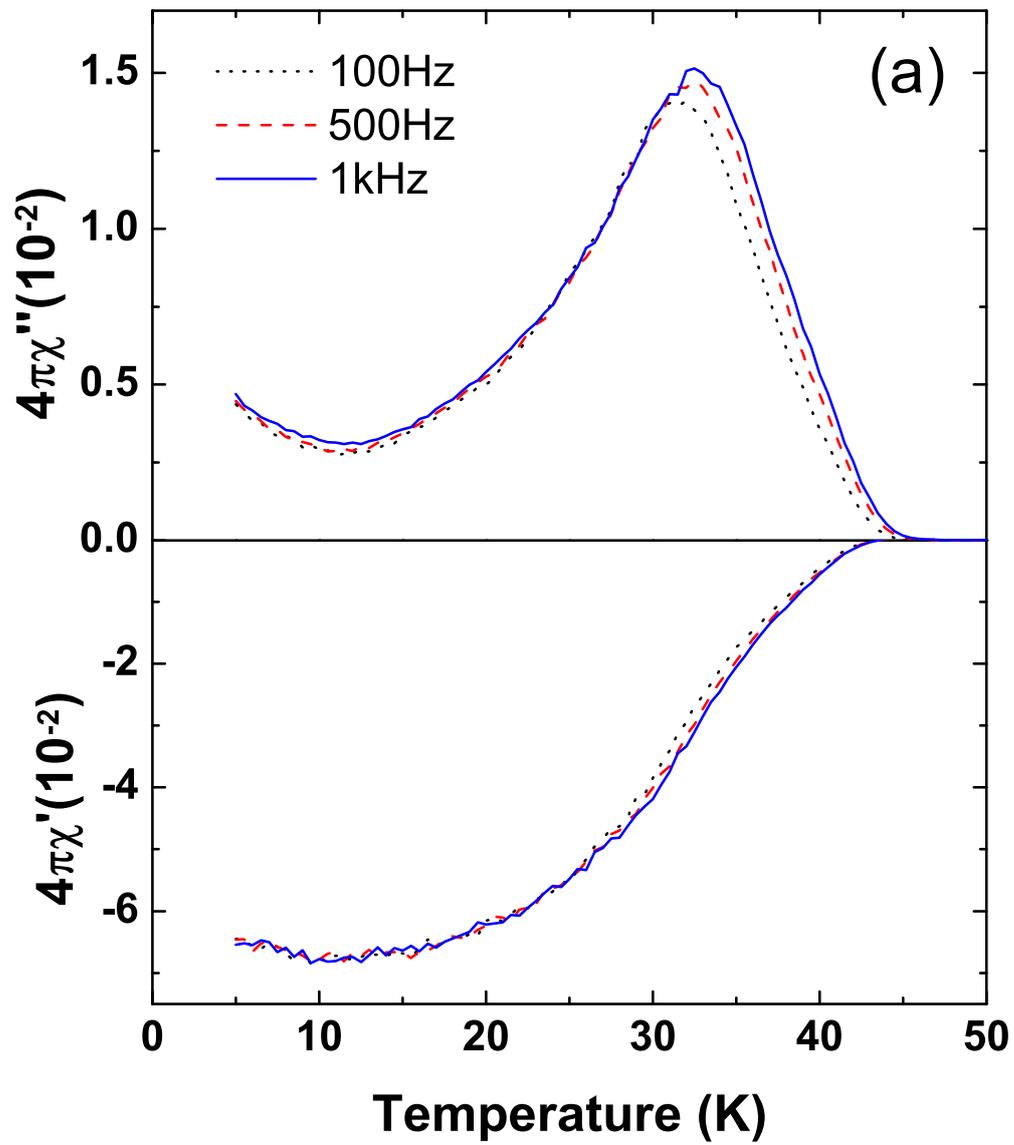

**Figure 4 (a) Y. Ding** *et al.*

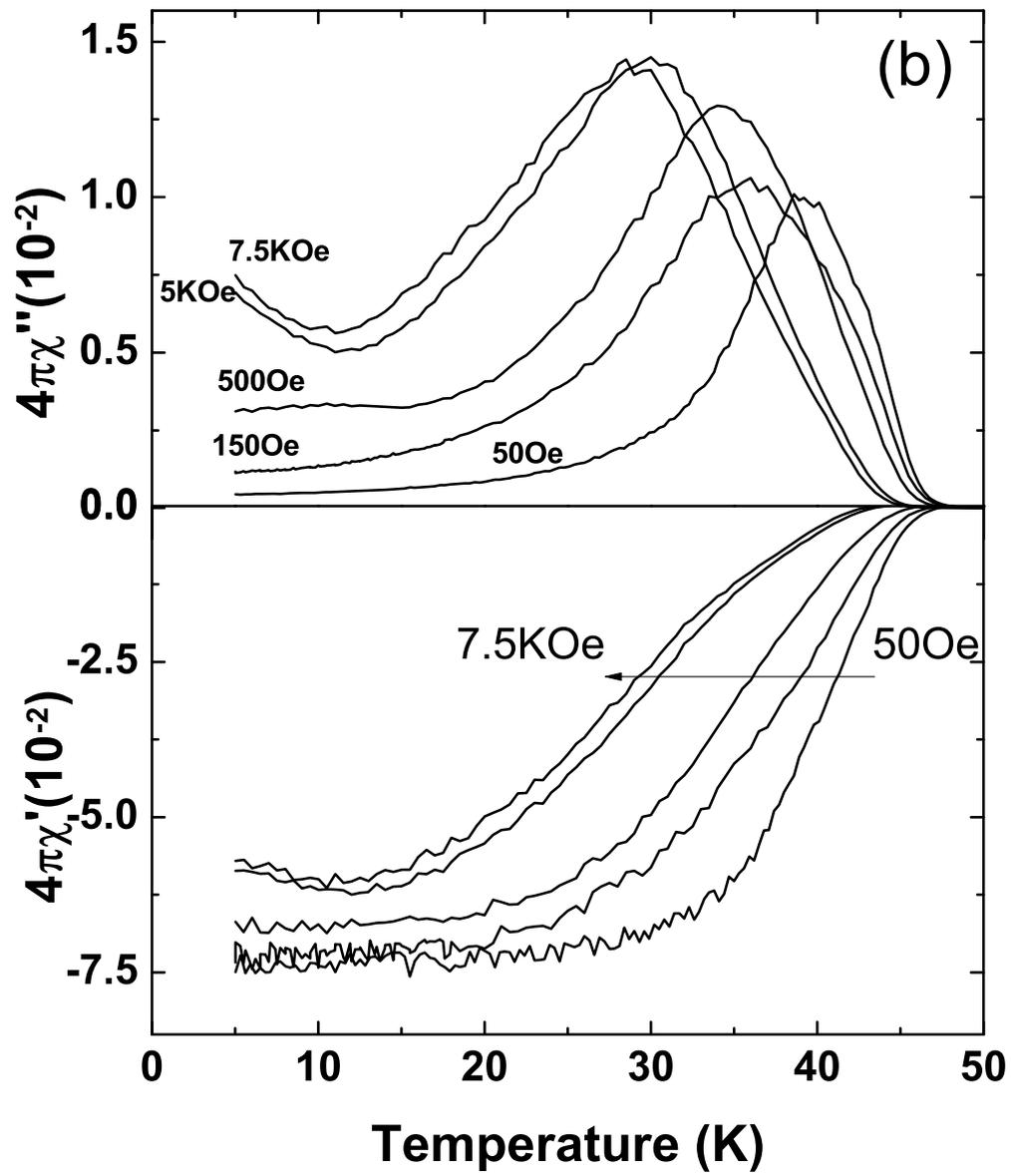

Figure 4 (b) Y. Ding *et al.*

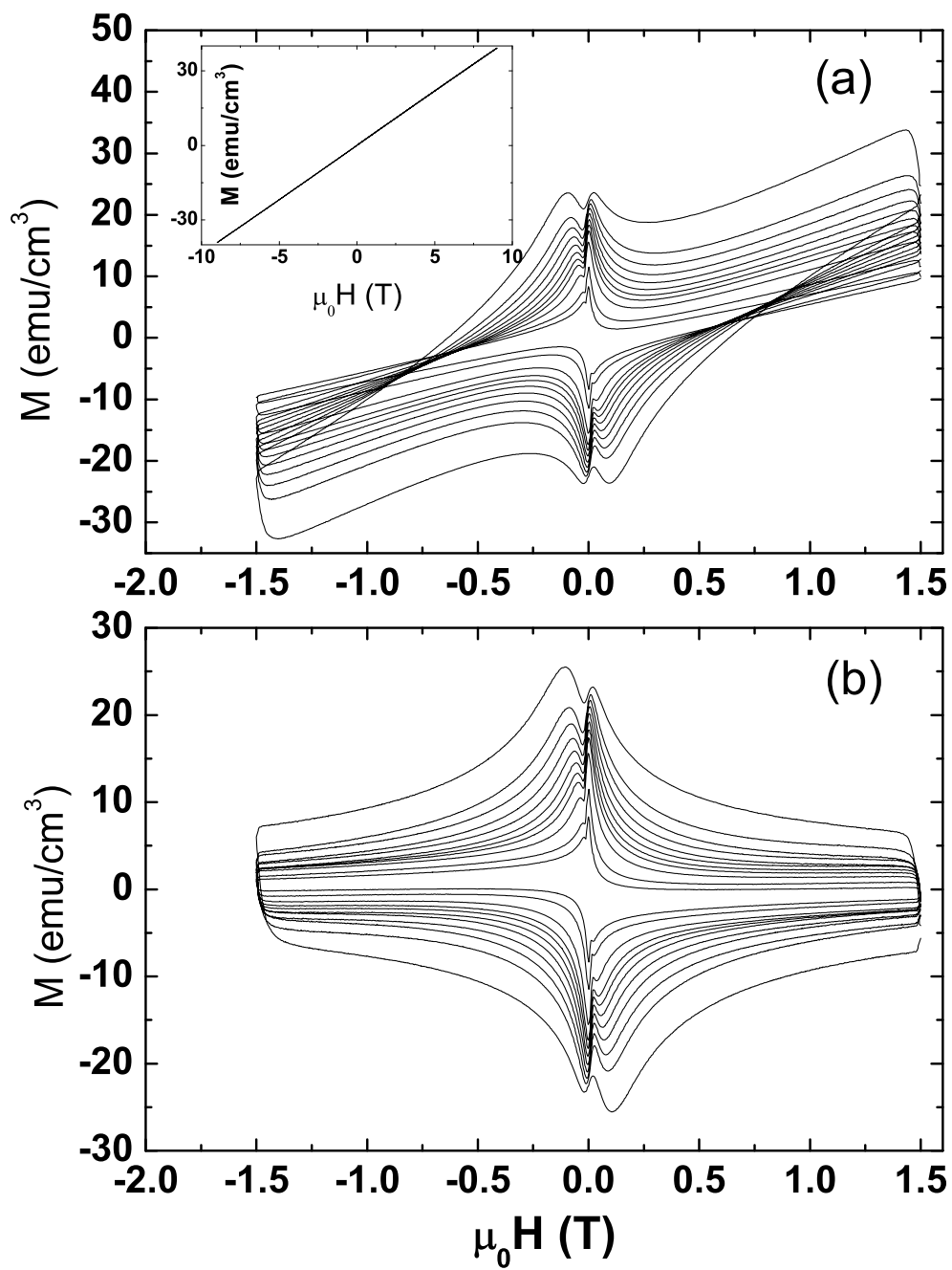

**Figure 5** Y. Ding *et al.*

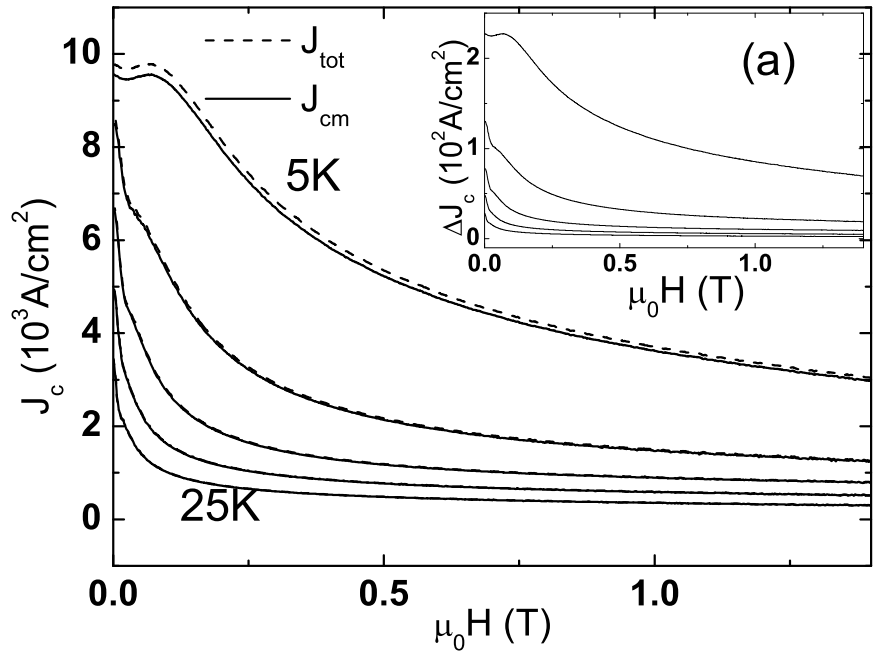

**Figure 6 (a) Y. Ding *et al.***

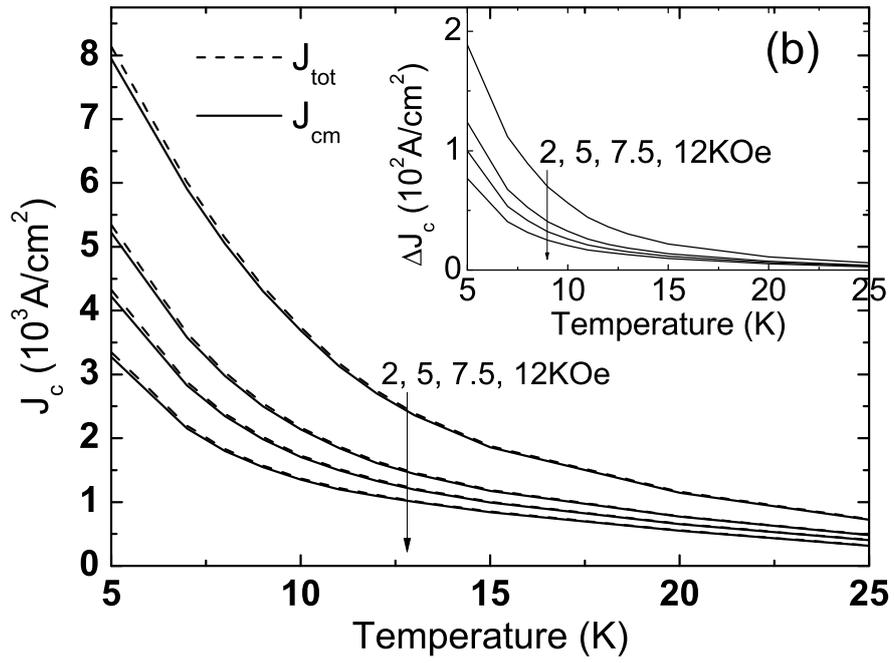

**Figure 6 (b) Y. Ding** *et al.*